\documentclass[conference]{IEEEtran}
\usepackage{cite}
\usepackage{amsmath,amssymb,amsfonts}
\usepackage{algorithm}
\usepackage{algpseudocode}
\usepackage{graphicx}
\usepackage{textcomp}
\usepackage{xcolor}
\usepackage{array}
\usepackage{caption}
\usepackage{tabularx}
\usepackage{bbding}
\usepackage{float}
\usepackage{subcaption}
\usepackage{booktabs}
\usepackage{makecell}
\usepackage[hidelinks]{hyperref}

\begin{document}
\title{Efficient Federated Intrusion Detection  in 5G ecosystem using optimized BERT-based model}

\author{\IEEEauthorblockN{Frédéric Adjewa}
\IEEEauthorblockA{\textit{LIST3N} \\
\textit{University of Technology of Troyes}\\
frederic.adjewa@utt.fr}

\and
\IEEEauthorblockN{Moez Esseghir}
\IEEEauthorblockA{\textit{LIST3N} \\
\textit{University of Technology of Troyes}\\
moez.esseghir@utt.fr}

\and
\IEEEauthorblockN{Leïla Merghem-Boulahia}
\IEEEauthorblockA{\textit{LIST3N} \\
\textit{University of Technology of Troyes}\\
leila.merghem\_boulahia@utt.fr}

}

\maketitle

\begin{center}
    \textbf{This work has been submitted to the IEEE for possible publication. Copyright may be transferred without notice, after which this version may no longer be accessible.}
\end{center}

\begin{abstract}
The fifth-generation (5G) offers advanced services, supporting applications such as intelligent transportation, connected healthcare, and smart cities within the Internet of Things (IoT). However, these advancements introduce significant security challenges, with increasingly sophisticated cyber-attacks. This paper proposes a robust intrusion detection system (IDS) using federated learning and large language models (LLMs). The core of our IDS is based on BERT, a transformer model adapted to identify malicious network flows. We modified this transformer to optimize performance on edge devices with limited resources. Experiments were conducted in both centralized and federated learning contexts. In the centralized setup, the model achieved an inference accuracy of 97.79\%. In a federated learning context, the model was trained across multiple devices using both IID (Independent and Identically Distributed) and non-IID data, based on various scenarios, ensuring data privacy and compliance with regulations. We also leveraged linear quantization to compress the model for deployment on edge devices. This reduction resulted in a slight decrease of 0.02\% in accuracy for a model size reduction of 28.74\%. The results underscore the viability of LLMs for deployment in IoT ecosystems, highlighting their ability to operate on devices with constrained computational and storage resources.
\end{abstract}


\section{Introduction}
In recent years, numerous technological revolutions have occurred in various fields, notably the emergence of new network architectures. The fifth generation (5G) is particularly noteworthy, offering faster and more scalable services. In addition to enhancing existing services, 5G enables new use cases such as intelligent transportation, connected healthcare, and smart cities, becoming a key element of the Internet of Things (IoT) \cite{yadav2022intrusion}. However, with these advancements, many security challenges arise from this heterogeneous and densely populated ecosystem. Cyber-attacks in 5G networks are becoming increasingly sophisticated and personalized, requiring the implementation of robust protection systems like intrusion detection systems (IDS). The primary goal of IDS is to identify various types of threats as early as possible to complement the shortcomings of firewalls. An intrusion can be seen as unauthorized activity that may cause damage to the system \cite{khraisat2019survey}. There are two main approaches to intrusion detection system: signature-based detection (SIDS) which uses predefined rules as a reference to identify malicious traffic. However, if an attack is not listed, this method is ineffective. On the other hand, anomaly-based detection system learns to recognize normal traffic flows and flags any deviation from the norm as suspicious \cite{rahman2020internet}. Machine learning and deep learning methods have made significant progress and are increasingly used to create new intrusion detection methods \cite{agrawal2022federated}. Another significant advancement is large language models which have shown their potential in many fields, including cybersecurity, where they have sometimes outperformed human experts \cite{tihanyi2024cybermetric}. However, it is essential to assess the risks associated with their use in sensitive tasks like computer security \cite{bhatt2024cyberseceval}.

In 5G environments, particularly in monitoring systems, personal information is used, so privacy concerns must be taken seriously. \cite{mcmahan2017communication} proposed federated learning, which allows training a global model collaboratively among multiple clients while keeping the data on their devices. Only the model parameters are exchanged, maintaining privacy and providing a reliable, scalable, and robust intrusion detection system. Federated learning can be mainly categorized into vertical federated learning, where user data do not share the same features but the observations are related, and horizontal federated learning, where users share the same observations space but with different features. Previous work in the field of intrusion detection relies on machine learning and deep learning approaches, such as random forests (RF), support vector machines (SVM), and gradient boosting (GB). Despite the effectiveness of these methods, they do not comprehensively provide a global security solution for deployment in 5G ecosystems, which are heterogeneous and densely populated. To overcome these limitations, we propose a federated anomaly-based Intrusion Detection System based on a LLM designed from BERT \cite{ferrag2024revolutionizing}, capable of deeply understanding the context of the data. The source code is available on Github.\footnote{https://github.com/freddyAdjh/Federated-security}

The main contributions of this paper are summarized as follows:

\begin{itemize}
\item A discussion of previous work on security in 5G architectures, with emphasis on intrusion detection systems.

\item A detailed procedure for implementing a lightweight and fast LLM for intrusion detection.

\item An Evaluation in both centralized and federated contexts of a large language model, demonstrating the effectiveness of the approach and its ability to ensure confidentiality and comply with regulations.

\item A Comparison of the model convergence when clients have IID versus non-IID data.

\item An experimental demonstration that the approach is suitable for resource-limited devices such as Raspberry Pi, with an inference time on CPU of 0.45s.

\end{itemize}

This paper is organized as follows. In the next section, we present related work on securing 5G networks, with a focus on anomaly-based intrusion detection systems (AIDS). Section III theoretically outlines the essential concepts of this paper. Then, Section IV presents the experimental results of implementing AIDS using a large language model (LLM) as the base model in different scenarios. Finally, we conclude.

\section{Related Work}
5G architectures bring numerous advantages through the services they offer. However, these connected services and applications remain susceptible to attacks aimed at compromising the network. Many studies have been conducted to propose solutions to address these security vulnerabilities.

\begin{figure}[ht!]
    \centering
    \includegraphics[width=0.65\linewidth]{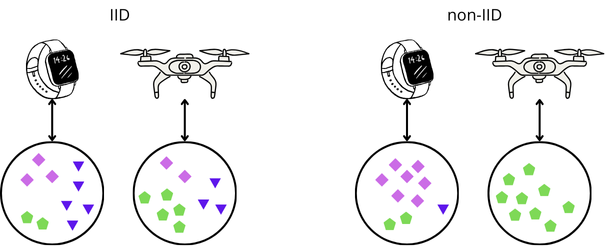}
    \caption{IID and Non-IID distribution}
    \label{dist}
\end{figure}

\cite{fang2017security} presented several solutions to secure the services offered by 5G, highlighting the numerous challenges faced by this technology. User privacy and data protection constitute a major issue that requires attention from researchers. \cite{ali2023security} demonstrated that user information could be easily accessible within the network, thereby exposing users to malicious actors.\cite{rahman2020internet} and \cite{khraisat2019survey} each analyzed intrusion detection systems (IDS), presenting the foundations, necessity, and implementation methods.
\cite{wei2021federated} proposed an architecture highlighting the deployment of IDS in a 5G ecosystem to provide a reliable security system at all levels.
To detect intrusions in the 5G network, \cite{yadav2022intrusion} proposed an intelligent intrusion detection system to identify attacks in IoT environments using deep learning methods.  \cite{zhang2023iot} evaluated machine learning approaches on IID and non-IID partitions. Fig. \ref{dist} shows how data are distributed in these different scenarios. The rapid evolution of technologies and services in 5G has motivated the implementation of federated learning, which enables learning while preserving data privacy.\cite{korba2023federated} presented the limitations of traditional machine learning and deep learning methods against unknown attacks and proposed the use of autoencoders to analyze the traffic in connected vehicle networks.
\cite{zhao2018federated} focused on the challenge of non-IID data, demonstrating its impact on model performance. They showed that the reduction in performance is linked to weight divergence caused by the nature of the data used. However, the proposed solution contradicts the principle of federated learning, although it seems to resolve the issue.
\cite{tihanyi2024cybermetric} and \cite{bhatt2024cyberseceval} discussed the involvement of large language models (LLMs) in cybersecurity, demonstrating their robustness but also the need to regulate their use to limit potential drawbacks. 
\cite{agrawal2022federated} conducted a study on the use of federated learning in 5G networks as a means of ensuring security while preserving confidentiality, emphasizing the importance of properly handling non-IID data. Based on these foundations, recent works have extended the use of federated architectures by incorporating LLMs. For instance, \cite{yan2024federa} proposed FeDeRA, an improvement of LoRA for use in a distributed ecosystem. Similarly, \cite{gao2024dlora} proposed an approach to distribute LoRA between the devices and the cloud.
However, all these works do not comprehensively cover all aspects related to deploying a security solution for 5G ecosystems. Table \ref{tab:comparTab} summarizes these aforementioned gaps. Our work aims to address these shortcomings by proposing a federated anomaly-based intrusion detection system that leverages the power of LLMs, ensuring robust security and user confidentiality.

\begin{table}[ht!]
    \centering
    \caption{Comparison with Recent Works.}
    \label{tab:comparTab}
    \begin{tabular}{|l|l|l|l|l|l|l|}
        \hline
          Ref. & \makecell{5G\\ Security} & \makecell{Data \\ Privacy} & \makecell{ Federated \\ Learning} & IDS & LLM &\makecell{ Model \\ Compression}\\ \hline

         \cite{fang2017security}&\checkmark & \checkmark& \XSolidBrush&\XSolidBrush &\XSolidBrush &\XSolidBrush\\ \hline
         \cite{ali2023security}&\checkmark & \checkmark& \XSolidBrush&\XSolidBrush &\XSolidBrush &\XSolidBrush\\ \hline
         \cite{yadav2022intrusion}&\checkmark & \XSolidBrush& \XSolidBrush&\checkmark &\XSolidBrush &\XSolidBrush \\ \hline
         \cite{zhang2023iot}&\XSolidBrush & \checkmark& \checkmark&\checkmark &\XSolidBrush &\XSolidBrush\\ \hline
         \cite{zhao2018federated}&\XSolidBrush & \checkmark& \checkmark&\XSolidBrush &\XSolidBrush &\XSolidBrush\\ \hline
         \cite{korba2023federated}&\XSolidBrush & \checkmark& \checkmark&\checkmark &\XSolidBrush &\XSolidBrush\\ \hline
         \cite{agrawal2022federated}&\XSolidBrush & \checkmark& \checkmark&\checkmark &\XSolidBrush &\XSolidBrush\\ \hline
        \cite{wei2021federated}&\checkmark & \checkmark& \checkmark&\XSolidBrush &\XSolidBrush &\XSolidBrush\\ \hline
        \cite{yan2024federa}&\XSolidBrush & \checkmark& \checkmark&\XSolidBrush &\checkmark &\XSolidBrush\\ \hline
        \cite{gao2024dlora}&\XSolidBrush & \checkmark& \checkmark&\XSolidBrush &\checkmark &\XSolidBrush\\ \hline
        \cite{ferrag2022edge}&\XSolidBrush & \checkmark& \checkmark&\XSolidBrush &\checkmark &\XSolidBrush\\ \hline
    \cite{ferrag2024revolutionizing}&\XSolidBrush & \checkmark& \XSolidBrush&\checkmark &\checkmark &\XSolidBrush\\ \hline
    \textit{Ours}&\checkmark & \checkmark& \checkmark&\checkmark &\checkmark & \checkmark\\ \hline
    \end{tabular}
\end{table}

\section{Methodology}
In this section, we will present the concept of Federated Learning. Following this, we will discuss realistic datasets used in the literature for the simulation of IDS. Then, we will introduce the model we used for detection. Next, we will cover data preparation, tokenization, and the application of linear quantization.

\subsection{Federated Learning}
Machine learning typically involves three main steps: data collection, where a certain amount of data is gathered on a central server; data processing, during which the collected data is cleaned to make it usable; and finally, the design of the learning model based on this dataset. However, data collection quickly raises privacy concerns and violates regulations from organizations such as the European Union's GDPR \cite{khalil2022federated}. To ensure user privacy and data protection, it is necessary to find a way to train AI models while keeping the data on the client device.

\begin{figure}[ht!]
\centering
\includegraphics[width=\linewidth]{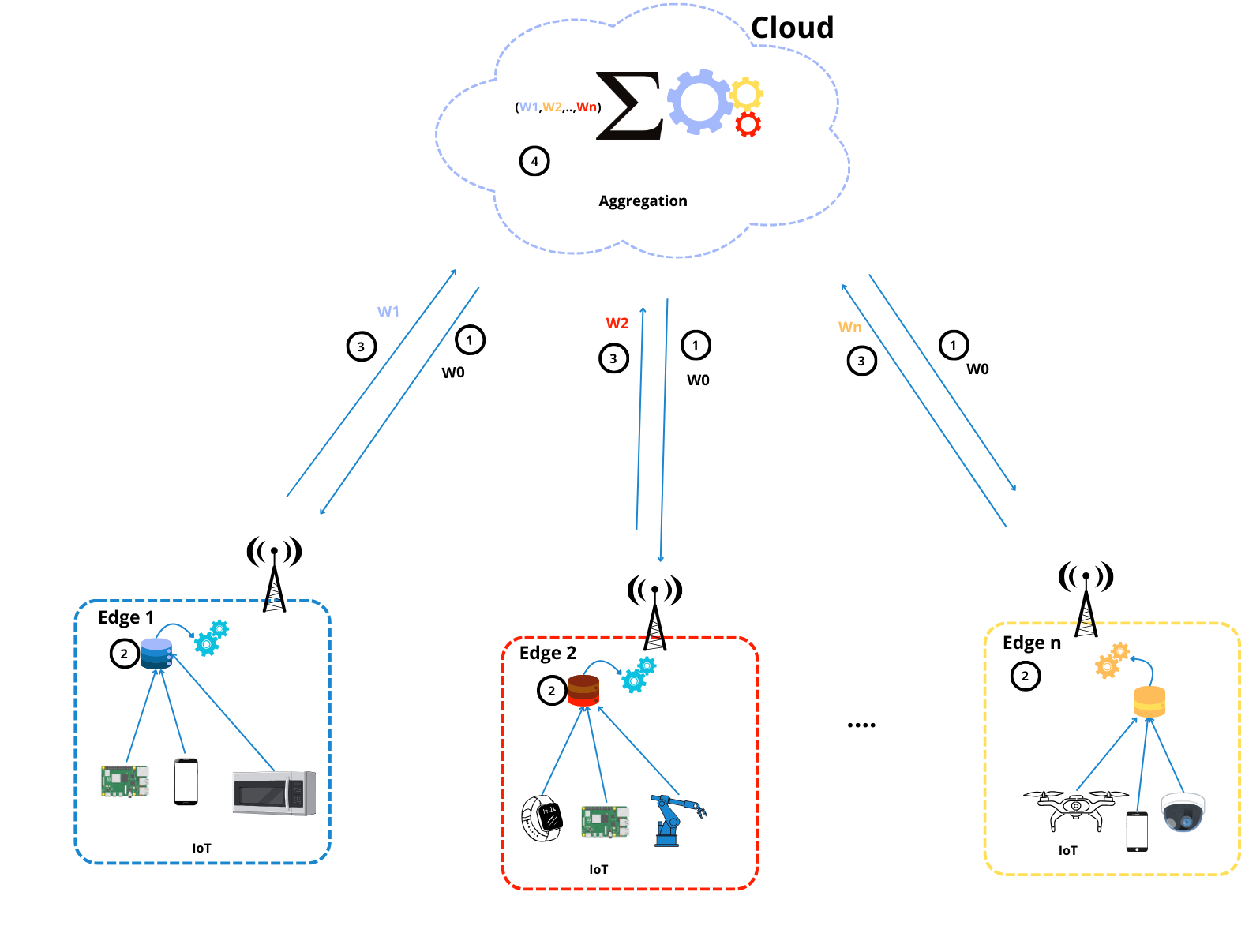}
\caption{Federated learning flow}
\label{FLflow}
\end{figure}

Federated learning addresses these issues by ensuring data privacy, pervasive intelligence in the environment, high flexibility, and good scalability \cite{agrawal2022federated}. As illustrated in Fig. \ref{FLflow}, federated learning is performed in four main steps:

\begin{enumerate}
\item Distribution of the global model to the edge devices.
\item Training of local models by the edge devices on their local data.
\item Transfer of weights from the edge devices to the aggregation server.
\item Aggregation of the received weights by the server using a specific aggregation strategy.
\end{enumerate}

\cite{mcmahan2017communication} proposed FedAvg, a robust algorithm that aggregates parameters received from edge devices while considering the proportion of data used for model training to weight this aggregation and obtain a reliable model. Several algorithms have since been proposed to optimize federated learning performance, whether on the client side or the server side \cite{ye2024openfedllm}. By using FedAvg in our experiments, we will demonstrate that the approach works with basic algorithm, suggesting that enhanced algorithms could provide even better results.

\subsection{Datasets}
Numerous studies present realistic datasets in the field of cybersecurity. Table \ref{tab:availds} compares some popular publicly available datasets that allow for the simulation of intrusion detection systems. However, some publicly available datasets do not realistically simulate certain scenarios that we encounter or may encounter. \cite{ferrag2022edge} proposed EdgeIIoTset, a comprehensive and realistic dataset ranked in the top 1\% of documents by Web of Science (WoS). For our experiments, we used this dataset in various proportions. The original dataset contains 61 features and 2,219,201 data points, but for speed considerations, in our study we used 30\% of the dataset. Table \ref{tab:classes} presents the distribution of the data used for federated experiments. To avoid biases in modeling, we did not use columns like "Timestamp" so that the model does not use the date as a feature and is thus able to detect an intrusion regardless of the time it occurs.

\begin{table}[ht!]
\centering
\caption{Summary of the classes}
\label{tab:classes}
\begin{tabular}{l|l|l|l}
\hline
\textbf{Classes}               & \textbf{Samples}  & \textbf{Train Set ($80\%$)} & \textbf{Test Set ($20\%$)} \\ \hline
Normal                & 484,693  & 387,754      & 96,939      \\ 
DDoS\_UDP              & 36,470   & 29,176       & 7,294       \\ 
DDoS\_ICMP             & 34,931   & 27,945       & 6,986       \\ 
SQL\_injection        & 15,361   & 12,289       & 3,072       \\ 
Password              & 15,046   & 12,037       & 3,009       \\ 
Vulnerability\_scanner & 15,033   & 12,026       & 3,007       \\ 
DDoS\_TCP              & 15,019   & 12,015       & 3,004       \\ 
DDoS\_HTTP             & 14,973   & 11,978       & 2,995       \\
\hline
\end{tabular}
\end{table}

\begin{table*}
    \centering
    \caption{Datasets available for simulating IDS}
    \begin{tabular}{|l|l|l|l|l|l|l|l|}
        \hline
         Dataset & Year & Number of Features & Number of Attacks & IoT & Label & Algorithm & Reference \\ \hline

         \textbf{NSL-KDD} & 2009 & 41 & 4 &\XSolidBrush & \checkmark & DNN, ANN, RF, KNN, SVM &\cite{tavallaee2009detailed} \\ \hline
          \textbf{UNSW-NB15} &  2015 & 49 & 9 &\XSolidBrush & \checkmark & DNN, ensemble methods, LSTM, SVM &\cite{moustafa2015unsw} \\ \hline
          \textbf{CIC-IDS2017} & 2017  & 80 & 8 & \XSolidBrush & \checkmark & KNN, RF, MLP, AdaBoost, Naive Bayes, ID3, QDA &\cite{sharafaldin2018toward} \\ \hline
          \textbf{BOT-IoT} & 2019 & 46 & 8 &\checkmark &\checkmark & SVM, LSTM, RNN &\cite{ferrag2022edge} \\ \hline
          \textbf{Edge-IIoTset} &2022  & 61 & 14  &\checkmark &\checkmark & DT, RF, KNN, SVM &\cite{ferrag2022edge} \\ \hline
    \end{tabular}
    \label{tab:availds}
\end{table*}

\subsection{Model design}
To construct the model, we used BERT \cite{devlin2018bert}, which is an encoder model. This model was adapted to our needs to create a model capable of predicting whether a flow is malicious or benign. We modified the base model by retaining only the first four layers of the encoder. The particularity of these lower layers is that they capture features related to syntax in the data they process. For each Transformer layer, we:

\begin{itemize}
    \item Used four encoder blocks instead of the twelve in the base model.
    \item Opted for a hidden size of 256 and intermediate size of 1024, from the rule $FFN=4H$\cite{devlin2018bert}.
    \item Considered the condition $H\%A=0$, leading us to use 4 attention heads instead of twelve.
\end{itemize}
Where $H$ represents the size of the hidden layer, $A$ the number of attention heads, $FFN$ the intermediate layer size and "$\%$" the modulo operator.
We then added a linear layer followed by a softmax layer at the output of the adapted BERT to perform the classification of the 15 states identified in the dataset.
As presented by \cite{ferrag2024revolutionizing}, the model named securityBERT has 11,174,415 parameters and a disk size of 42.63 MB. BERT base is 420MB, so these modifications led to a model size reduction of 89,85\%.

\subsection{Data Preparation}
The dataset contains numerical and string entries, which BERT cannot directly understand. Therefore, it was necessary to transform these data types to make them comprehensible to the model. First, we concatenated all the columns from each row of the dataset and then used a hash function to transform them, thereby improving privacy.

\subsection{Tokenization}

It was crucial to build a vocabulary allowing the model to understand the data processed. \cite{wang2020neural} proposed the Byte-Level Byte-Pair Encoder (BBPE), which enabled us to build this vocabulary by combining two bytes at a time. This method optimally manages the diversity of data and improves the model's ability to interpret nuances in the encoded flows.
We ensure that the data processed locally are encrypted and the parameters transmitted are from these transformed data and not from the raw data.

\subsection{Linear quantization}
Linear quantization is a technique used to reduce the memory footprint of any model by lowering the precision requirements for the weights and activations \cite{krishnamoorthi2018quantizing}. We employed per-channel linear quantization, where each channel of the weight matrix is quantized independently. This method allows for finer granularity and often leads to better performance compared to per-tensor quantization, where the entire matrix is quantized using a single scale and zero-point.

Due to its simplicity and efficiency, we used post-training quantization to reduce our model size. This approach is particularly beneficial for resource-limited devices, as smaller and more efficient models are crucial in such environments. 

\section{EXPERIMENTATIONS AND DISCUSSION}
In this section, we will present the experimental results obtained with securityBERT in a centralized context to demonstrate the model's effectiveness. Then, we will discuss the results in a federated context where securityBERT was trained collaboratively among edge devices and quantized to detect intrusions. Finally, we compare the results to highlight the model's performance in both contexts.

\subsection{Simulation Setup}

\begin{table}[ht!]
\centering
\caption{Hyperparameters used for the base model configuration}
\label{tab:baseconf}
\begin{tabular}{|l|l|}
\hline
Parameter & Value \\ \hline
Number of encoder layers & 4 \\ \hline
Number of attention heads (A) & 4 \\ \hline
Hidden size dimension (H)& 256  \\ \hline
Intermediate layer size (FFN) & 1024 \\ \hline
Sequence length & 512 \\ \hline
Number of tokens & 5000 \\ \hline
Batch size & 32 \\ \hline
Learning rate & 5e-5 \\ \hline
Weight decay & 1e-3 \\ \hline
Number of epochs & 4 \\ \hline
Dropout & 0.1 \\ \hline
$\alpha$ (concentration) & 0.07 \\ \hline
\end{tabular}
\end{table}

The simulations we conducted were executed using Python 3.10.14 and PyTorch. The computational resources, including a Quadro RTX 6000 GPU with 22.5 GB of memory and Intel(R) Xeon(R) Gold 6128 CPU @ 3.40GHz, were provided by the regional computing center ROMEO \cite{Romeo}. In the federated setup, we simulated heterogeneous data handled by each client using Latent Dirichlet Allocation \cite{onan2016lda} to reflect real-world scenarios. When the concentration parameter tends to infinity, the data are IID; when it tends to zero, the data tend to non-IID. Resources were evenly distributed among the edge devices. Table \ref{tab:baseconf} outlines all the hyperparameters we used to construct the base model, which was employed for both centralized and federated evaluations.

\subsection{IDS Based Centralized Learning}

\begin{figure}[ht!]
    \centering
    \includegraphics[width=0.75\linewidth]{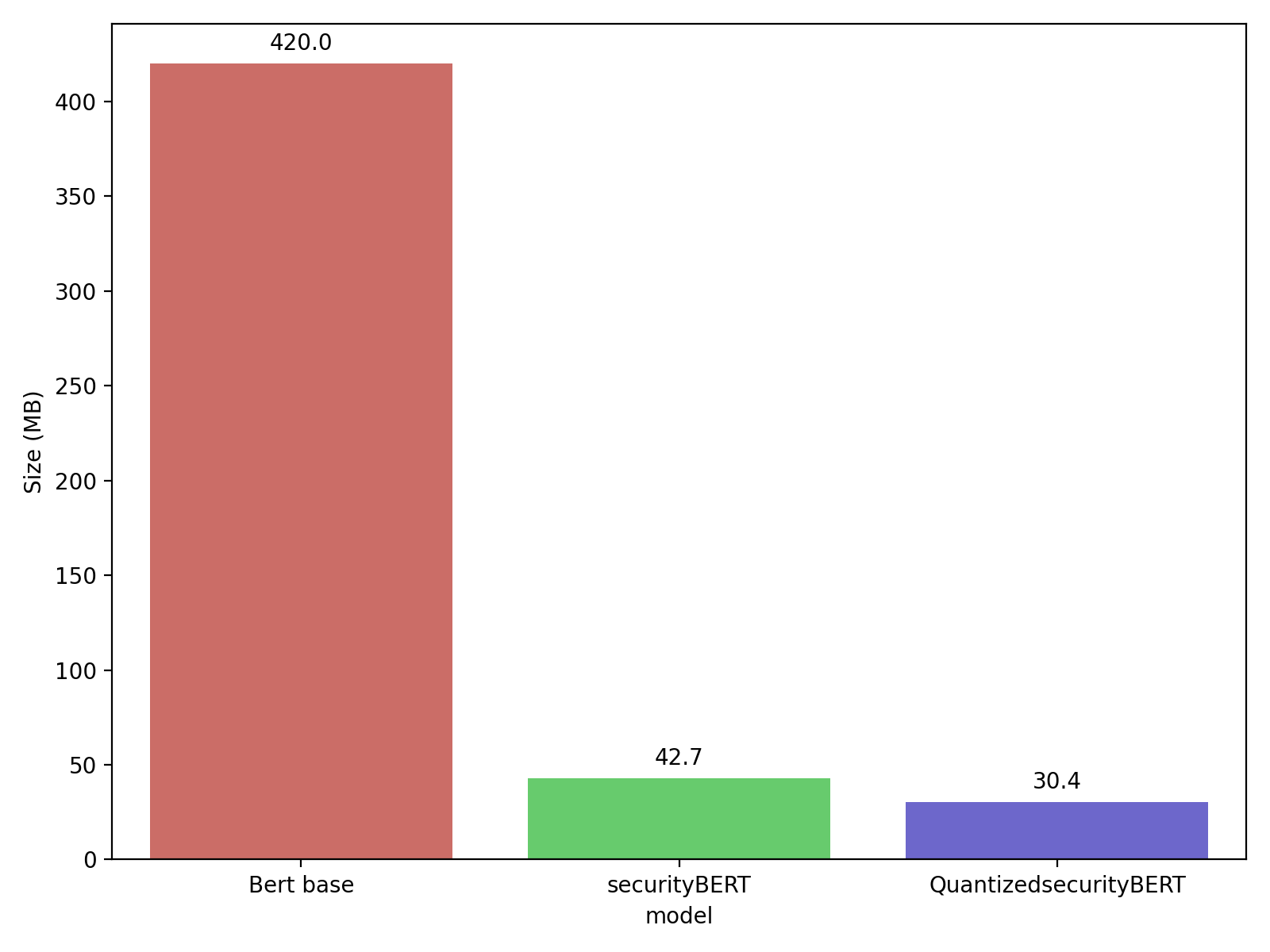}
\caption{Comparison of model sizes: initial, reduced, and quantized}

    \label{fig:model sizes}
\end{figure}

Fig. \ref{fig:model sizes} illustrates the difference in model size after the modifications. Initially, the base model weighed around 420MB, which can be considered somewhat heavy. After retaining only the first four encoder layers, we achieved a reduction of approximately 89.85\%, which is quite significant, especially since we aim to deploy the model on edge devices. However, there is potential for further reduction. As mentioned earlier, we applied linear quantization, resulting in an overall model size reduction of 92.76\%. This makes the model easily deployable on resource-constrained devices.

\begin{figure}[ht!]
    \centering
    \includegraphics[width=0.75\linewidth]{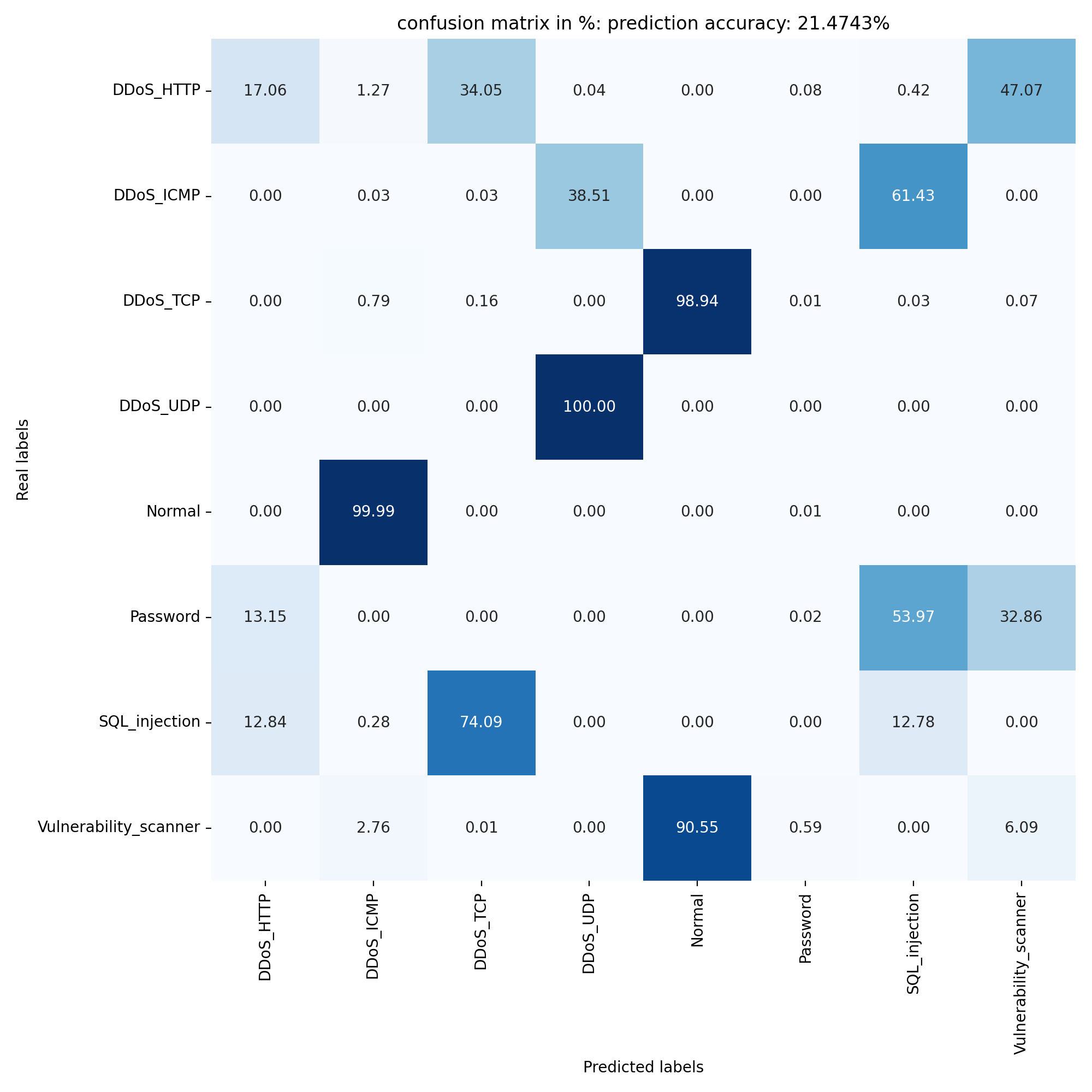}
    \caption{Confusion matrix of the model without training on network flow}
    \label{fig:confusion matrix without}
\end{figure}

After modifying the BERT base model, we evaluated whether this model possesses knowledge about intrusion detection, specifically in relation to the data we are working with. The confusion matrix shown in Fig. \ref{fig:confusion matrix without} demonstrates that the model lacks specific knowledge relevant to our downstream task. The classification results are poor, stemming from the foundational knowledge acquired during the model's pre-training phase.

\begin{figure}[ht!]
\centering

\begin{subfigure}[b]{0.24\textwidth}
\centering
\includegraphics[width=\textwidth]{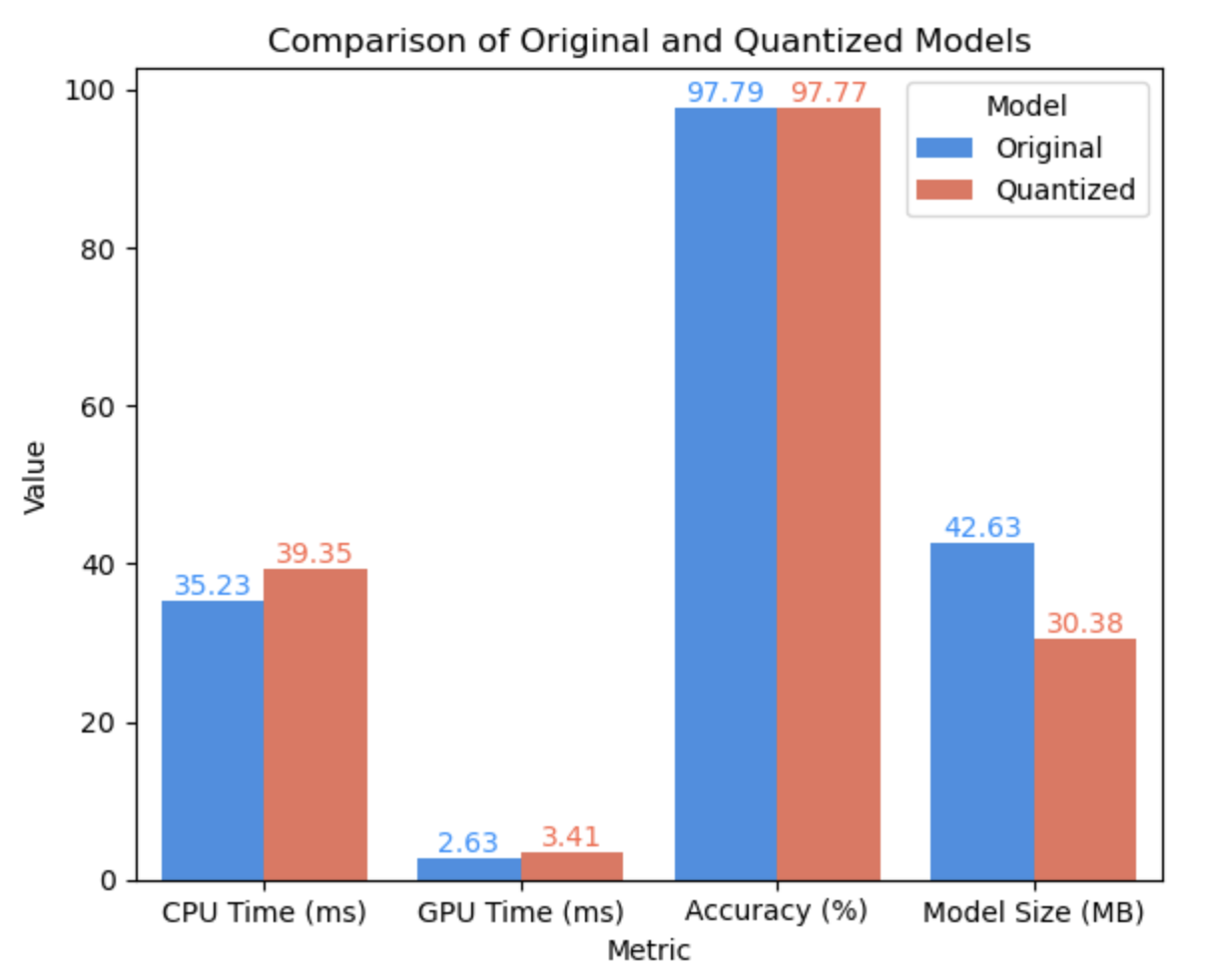}
\caption{Original vs. Quantized model performances}
\label{quant}
\end{subfigure}
\hfill
\begin{subfigure}[b]{0.24\textwidth}
\centering
\includegraphics[width=1.15\textwidth]{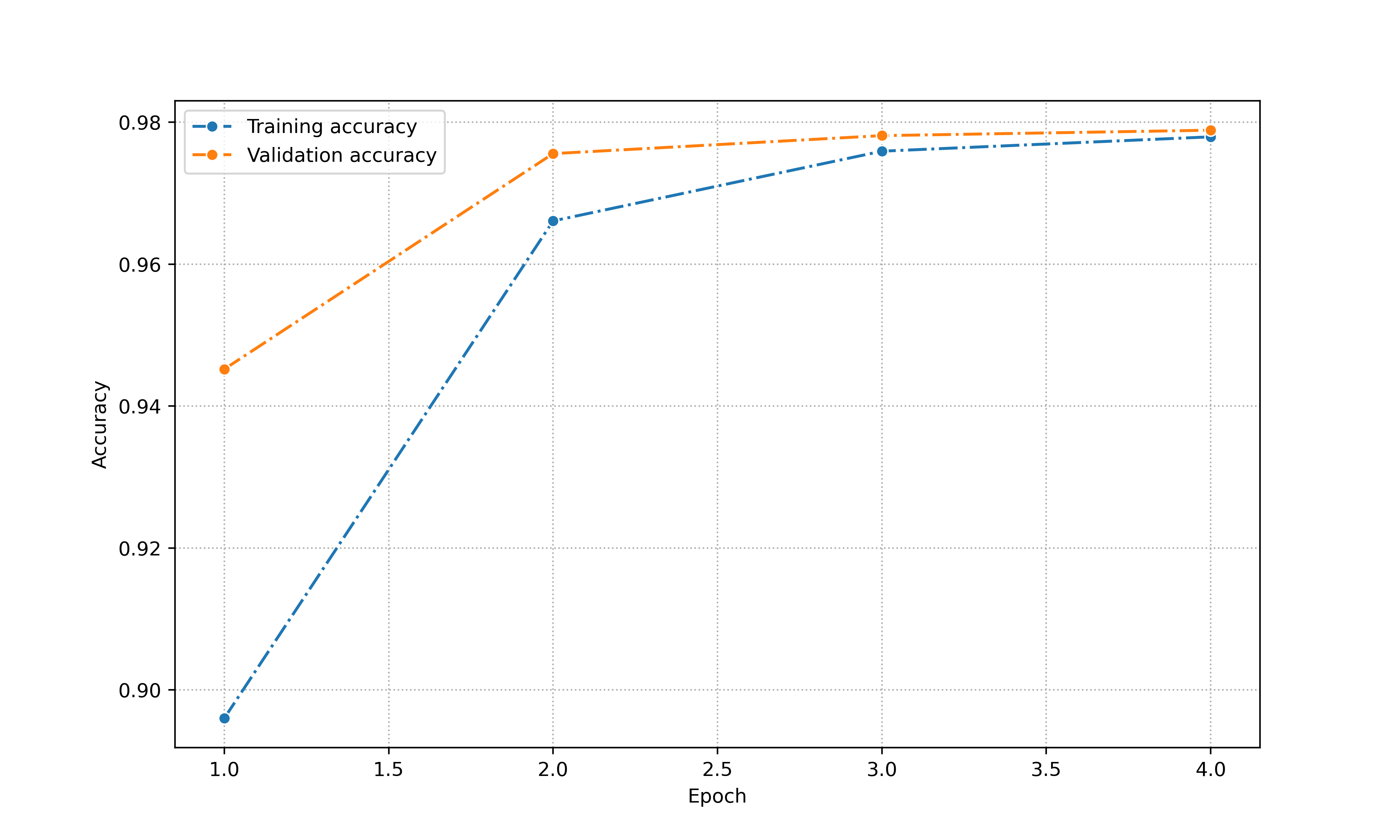}
\caption{Training and Evaluation accuracy}
\label{fig:accTraining}
\end{subfigure}
\hfill
\begin{subfigure}[b]{0.24\textwidth}
\centering
\includegraphics[width=1.15\textwidth]{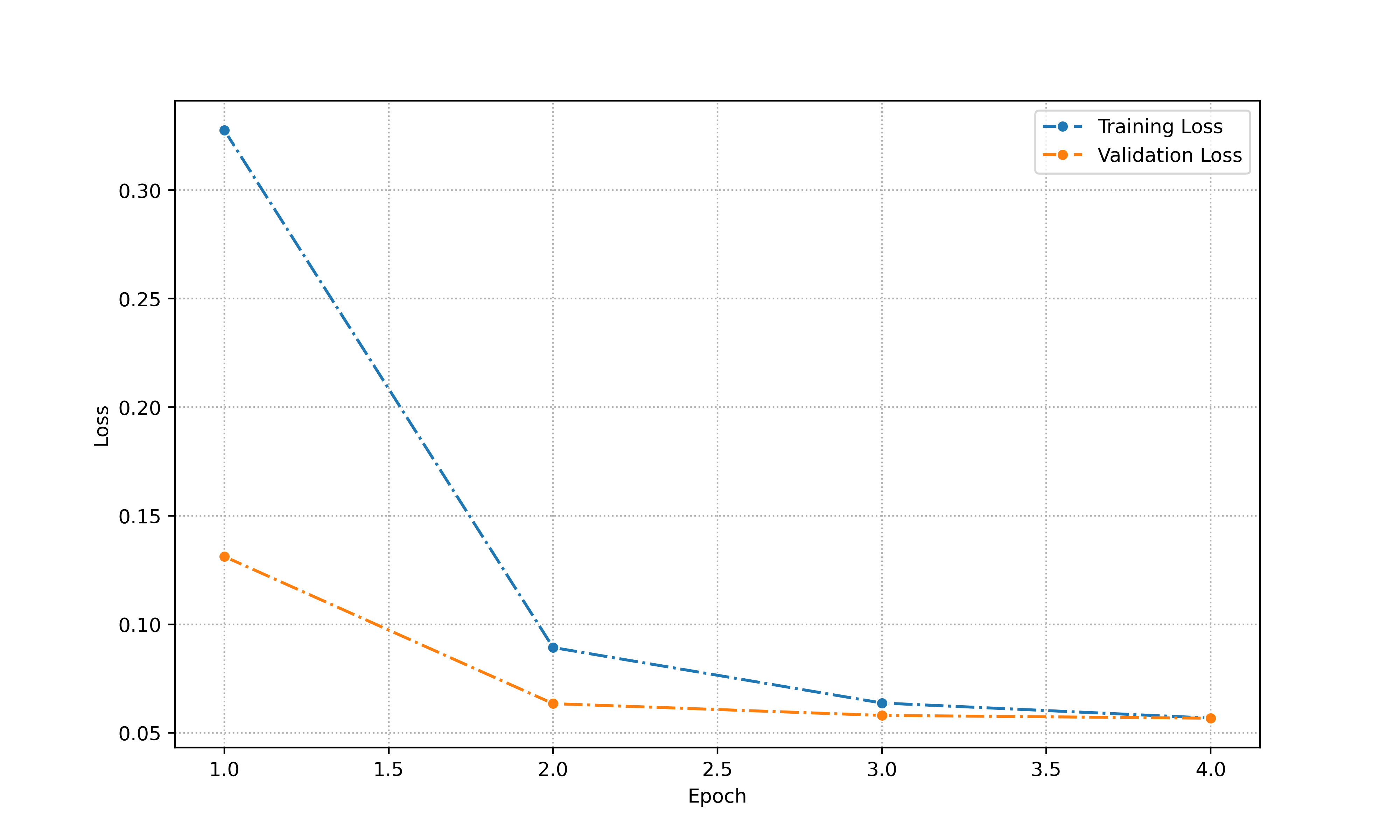}
\caption{Training and Evaluation loss}
\label{fig:lossTraining}
\end{subfigure}

\caption{securityBERT's accuracy and loss metrics over epochs}
\end{figure}

We trained the model to detect 14 different types of attacks, achieving an accuracy of 97.79\%. This result is very close to the 98.2\% obtained in \cite{ferrag2024revolutionizing}. The slight difference, however, would be due to the reduction in data volume, as mentioned above. Fig. \ref{quant} shows that the original model has an inference time of 35.23 ms on an Intel(R) Xeon(R) CPU, and 2.63 ms on a GPU. The same figure shows that by compressing the model by approximately 28.74\%, we experience a slight loss in accuracy, around 0.02\%, which is not significant and does not noticeably degrade the overall centralized performance. Figures \ref{fig:accTraining} and \ref{fig:lossTraining} depict centralized performance, and we can observe the beginning of overfitting by the 3rd epoch.

\subsection{IDS Based Federated Learning}

\begin{figure}[ht!]
    \centering
    \begin{subfigure}[b]{0.24\textwidth} 
        \centering
        \includegraphics[width=\linewidth]{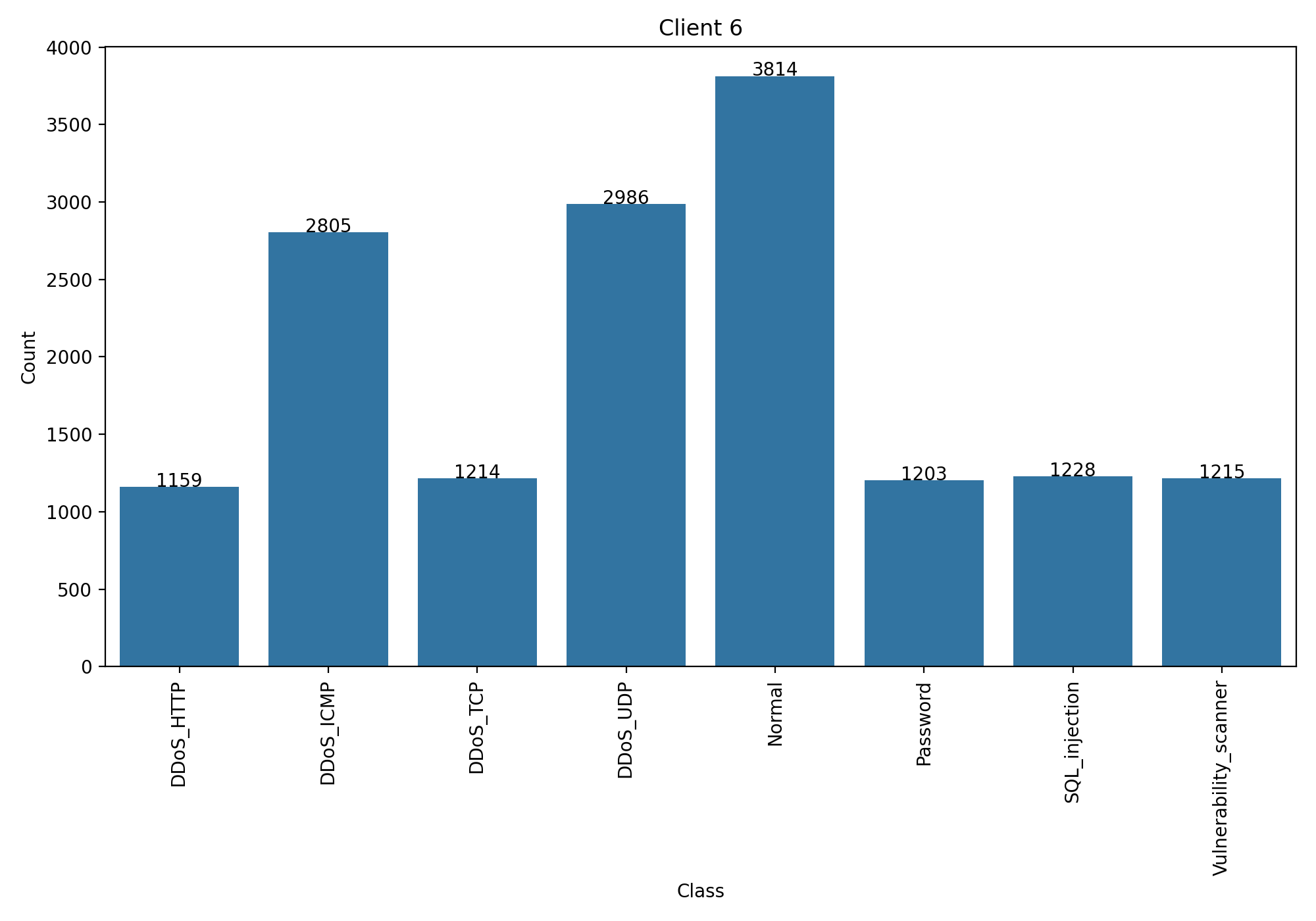}
        \caption{Client 6 IID}
        \label{fig:client 6}
    \end{subfigure}
    \hfill 
    \begin{subfigure}[b]{0.24\textwidth}
        \centering
        \includegraphics[width=\linewidth]{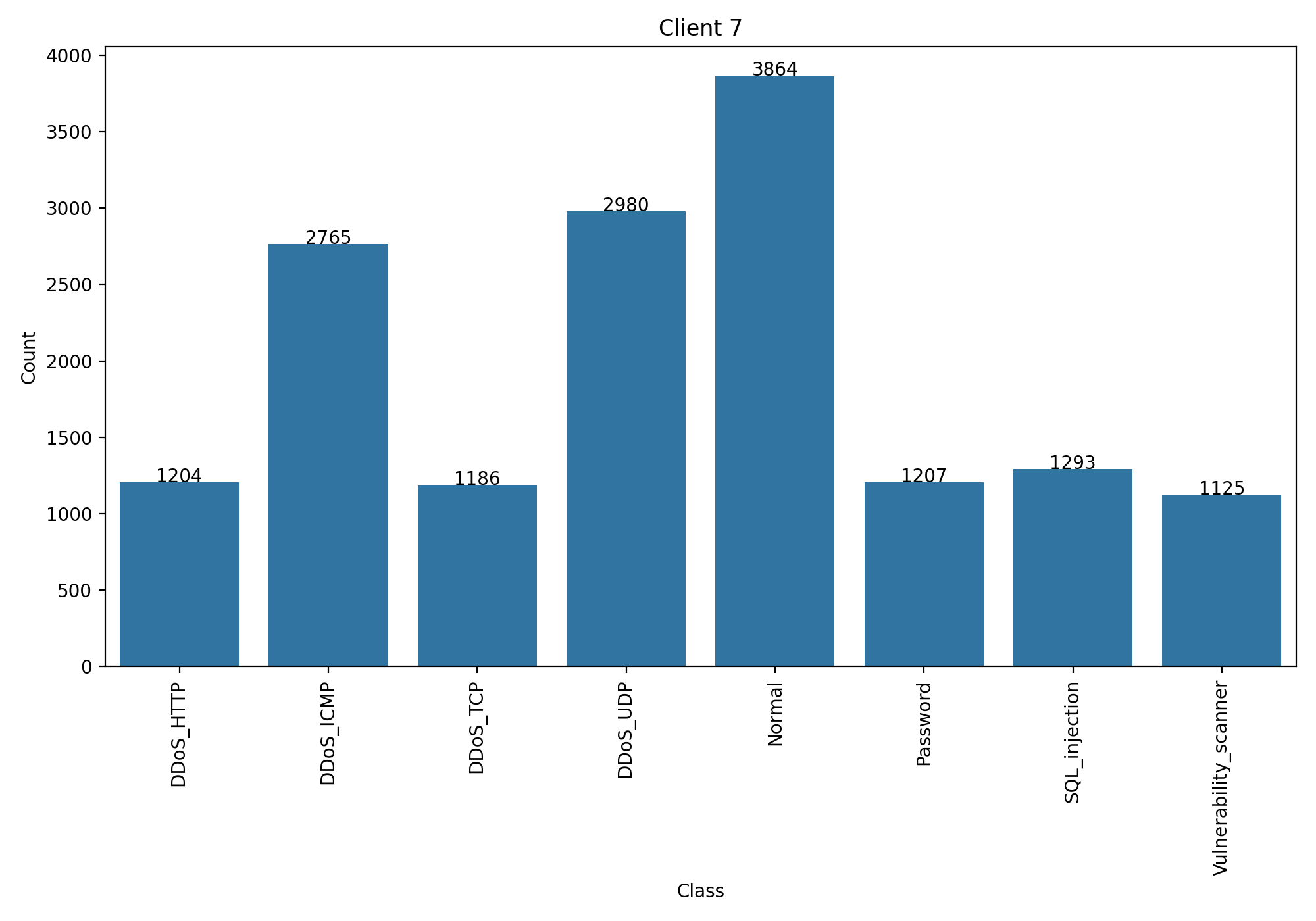}
        \caption{Client 7 IID}
        \label{fig:client§}
    \end{subfigure}
    \hfill 
    \begin{subfigure}[b]{0.24\textwidth}
        \centering
        \includegraphics[width=\linewidth]{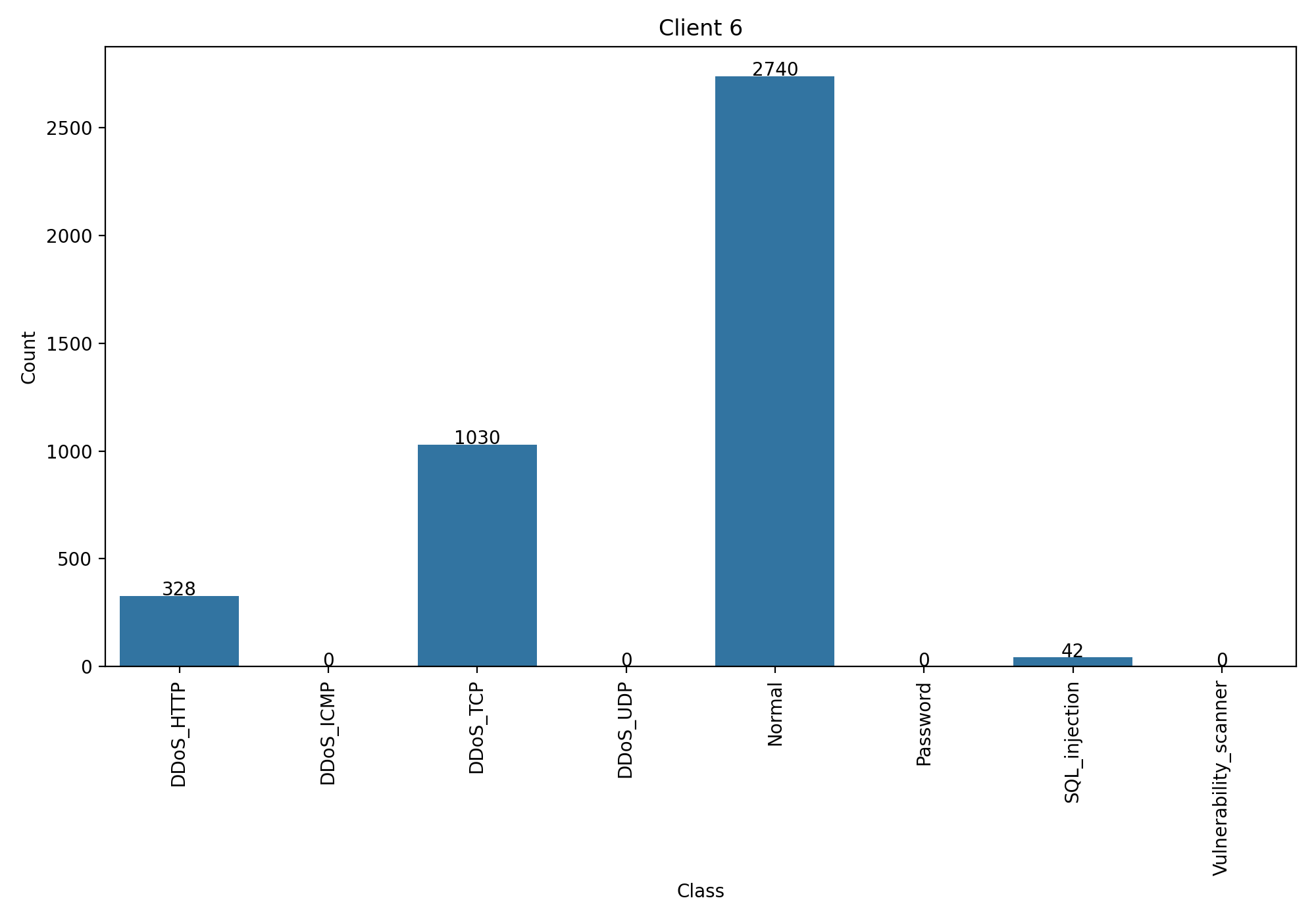}
        \caption{Client 6 non-IID}
        \label{fig:client6n}
    \end{subfigure}
    \hfill
    \begin{subfigure}[b]{0.24\textwidth}
        \centering
        \includegraphics[width=\linewidth]{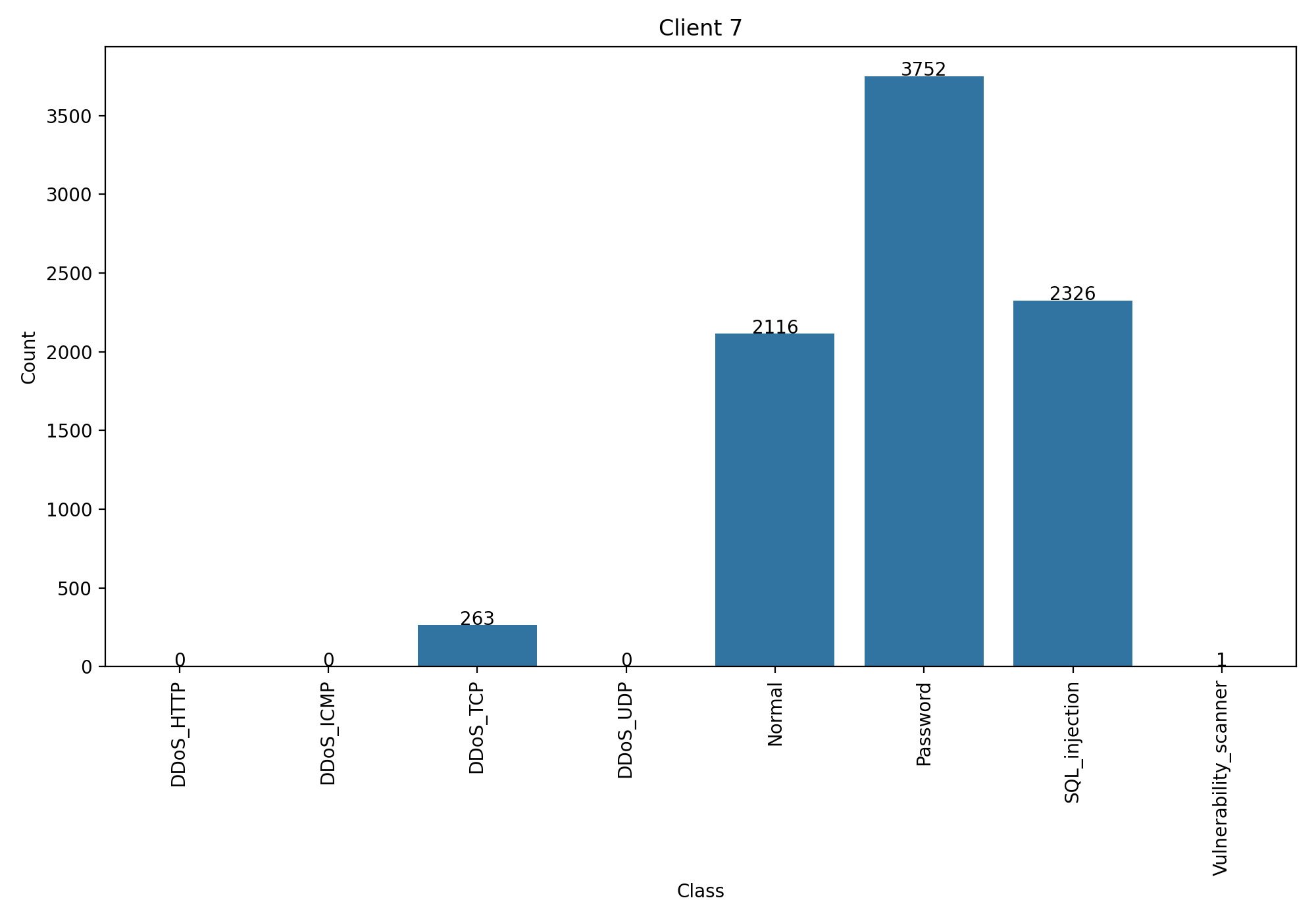}
        \caption{Client 7 non-IID}
        \label{fig:client7n}
    \end{subfigure}
    
    \caption{Data Distribution for two randomly selected clients in both IID and non-IID configuration}
    \label{fig:clients distribution}
\end{figure}

We simulated a federated environment where local models were trained on simulated edge devices, each having 2 CPU cores, and periodically communicated with a central server for model aggregation. As demonstrated in centralized learning, the model performs well in detecting 14 types of attacks, regardless of the total number of samples for each of them. In this scenario, we chose to focus on the attacks with the highest number of samples. Throughout the rest of the simulation, we address eight classes, detailed in Table \ref{tab:classes}.

Fig. \ref{fig:clients distribution} shows the distribution of two random clients from the ten sampled in both IID and non-IID simulations. When the data are IID, all clients possess all classes in approximately the same proportions. This can be interpreted as the network flow through them being similar. However, this scenario is not realistic, which is why we choose to work with non-IID data, utilizing a very low concentration parameter (0.07) to achieve greater realism. With non-IID data, we observe that, unlike in the IID case, clients do not exhibit similar distributions; some clients may lack certain classes, and even when they share classes, the proportions vary. This indicates that the flow through the clients is distinctly different, thus rendering it more realistic.

\begin{figure*}[ht!]
    \centering
    
    \begin{subfigure}[b]{0.24\textwidth} 
        \centering
        \includegraphics[width=\linewidth]{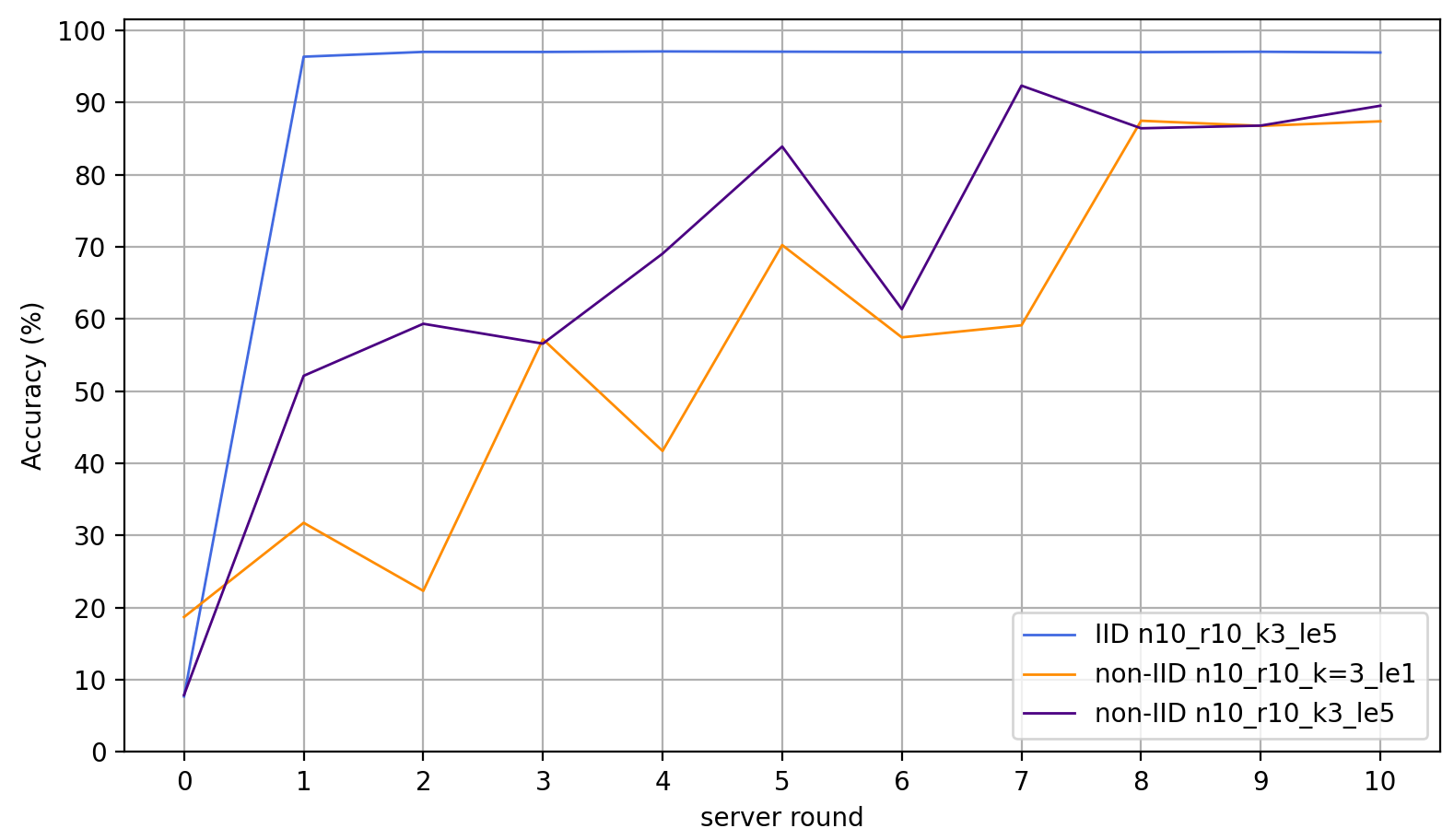}
        \caption{Performances with $K=3$}
        \label{fig:k3}
    \end{subfigure}
    \hfill 
    \begin{subfigure}[b]{0.24\textwidth}
        \centering
        \includegraphics[width=\linewidth]{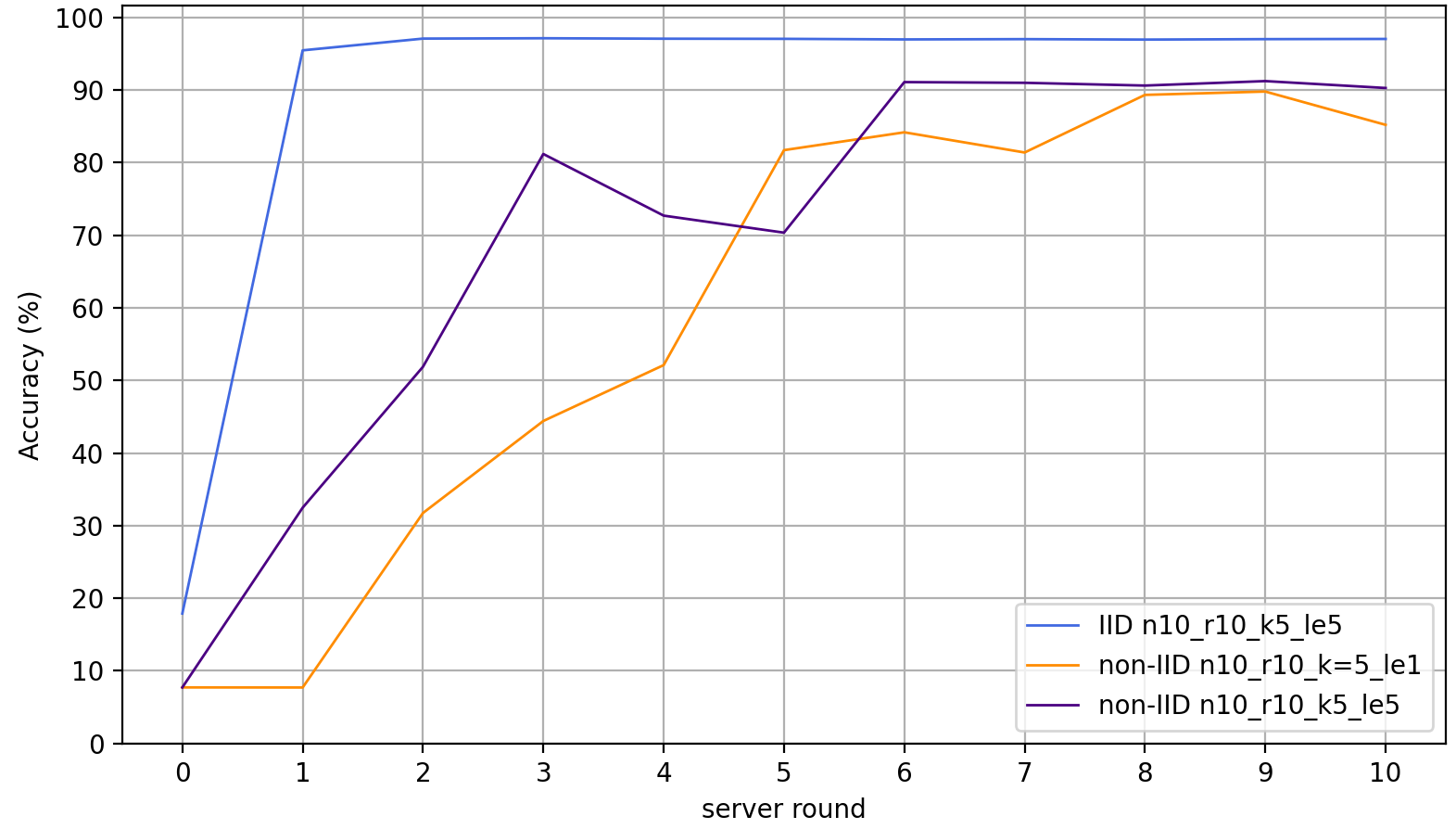}
        \caption{Performances with $K=5$}
        \label{fig:k5}
    \end{subfigure}
    \hfill 
    \begin{subfigure}[b]{0.24\textwidth}
        \centering
        \includegraphics[width=\linewidth]{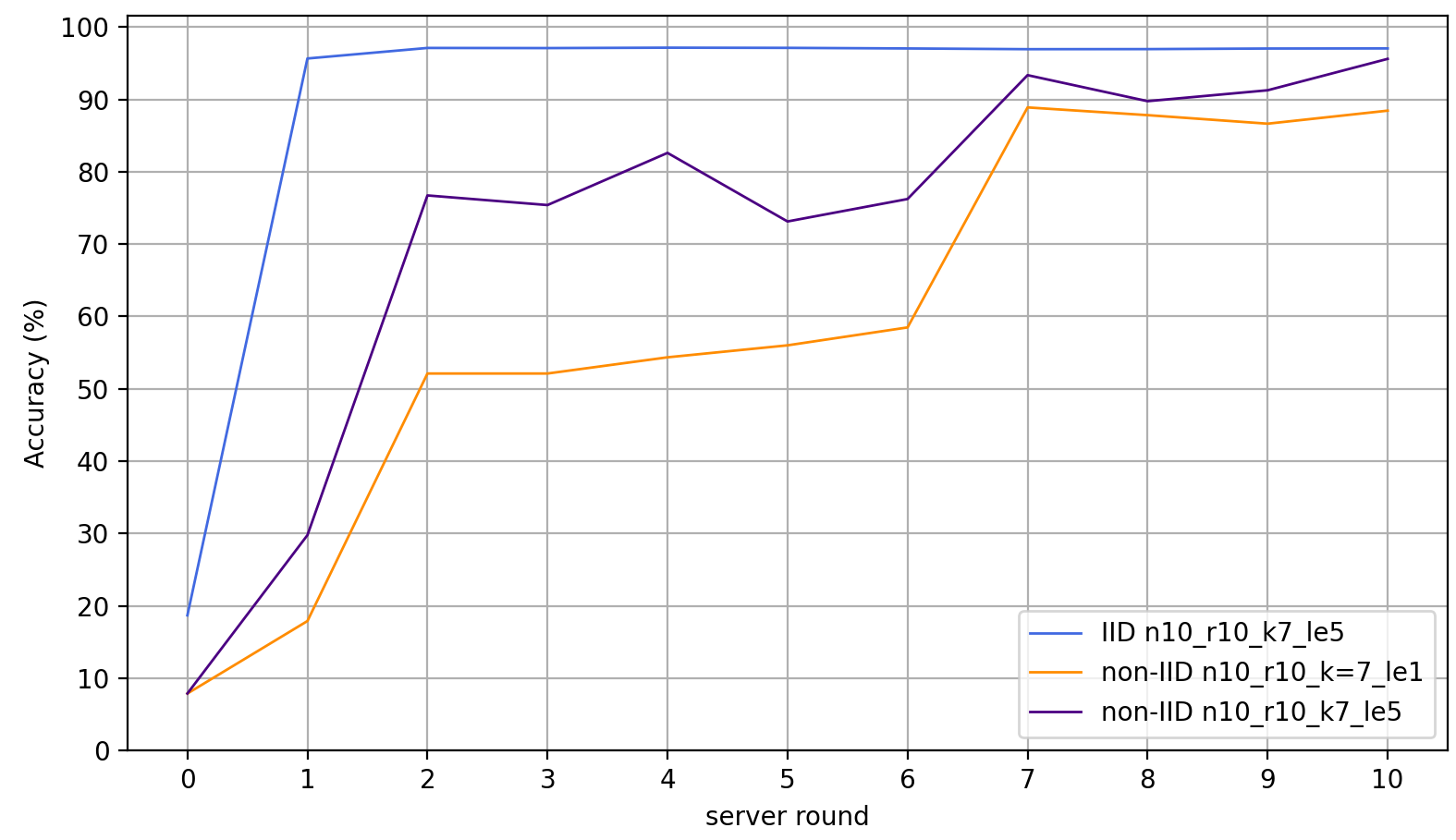}
        \caption{Performances with $K=7$}
        \label{fig:k7}
    \end{subfigure}
    \hfill
    \begin{subfigure}[b]{0.24\textwidth}
        \centering
        \includegraphics[width=\linewidth]{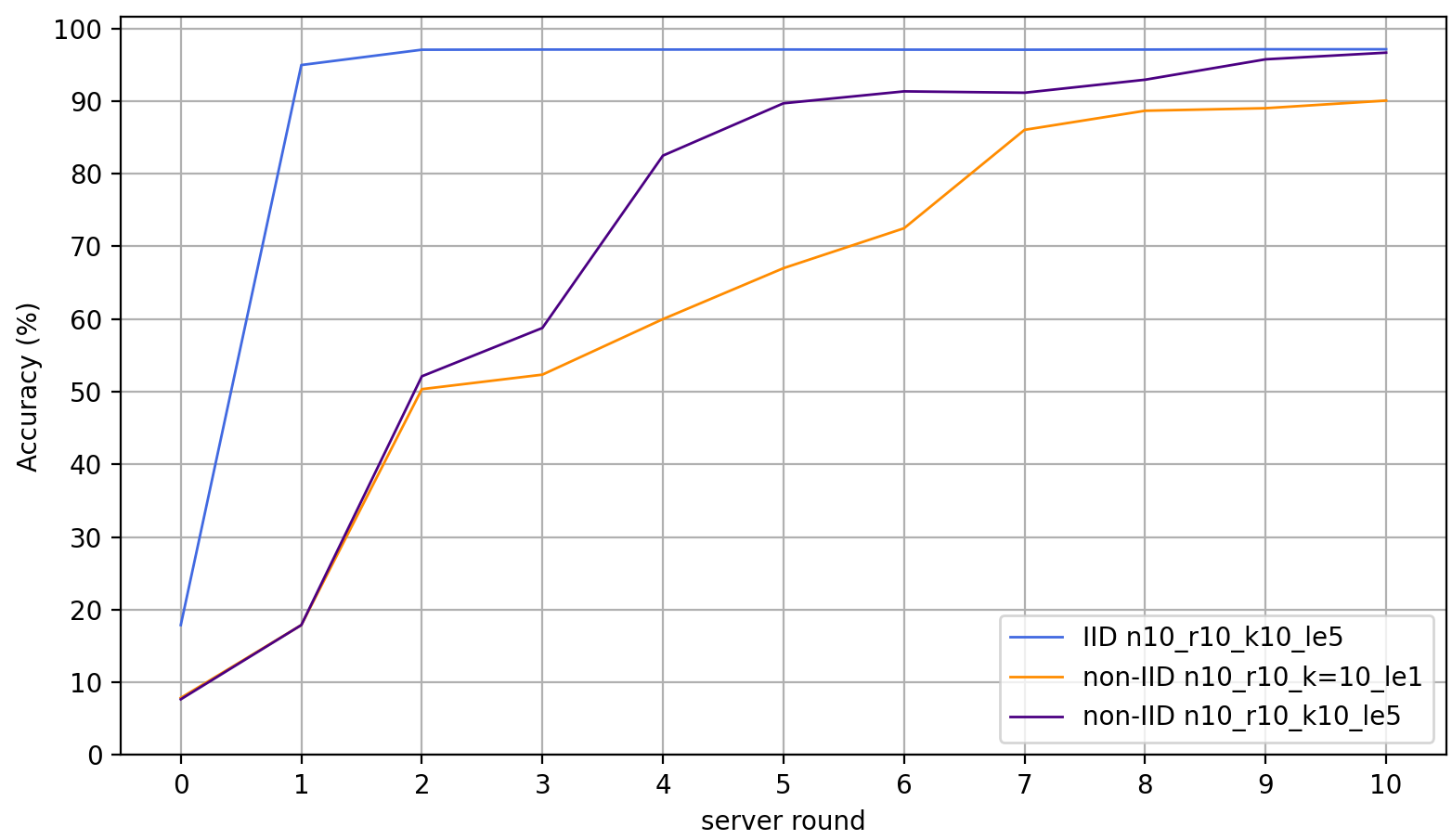}
        \caption{Performances with $K=10$}
        \label{fig:k10}
    \end{subfigure}
    
    \caption {Comparative analysis of detection model performance on IID and Non-IID data with different parameters}
    \label{fig:parameter-comparison}
\end{figure*}

Fig. \ref{fig:parameter-comparison} illustrates the variation in the global model's accuracy over epochs with a variable set of client and with different number of local epochs. At epoch 0, the aggregation server evaluates the model initialized with random parameters. The figure shows that the model struggles to converge with non-IID data, likely due to the significant differences between the clients' data sets. As a result, at certain epochs, the accuracy of the global model decreases after aggregating newly learned parameters. 

When $K$ is large, the model experiences fewer convergence issues and tends to achieve performance closer to that of IID data, with the model reaching approximately 95\% to 97\% accuracy by the third epoch. Observing the orange curve, which represents clients training locally for only one epoch, we see that the model is highly inconsistent. This suggests that the clients do not have sufficient time to share meaningful information with the aggregation server, leading to the global model struggling to converge and requiring significantly more time to achieve satisfactory accuracy. 

In contrast, the indigo curve represents the performance of the global model when clients train for more epochs locally. With longer local training, clients have enough time to transmit more meaningful information about their data to the server, resulting in fewer accuracy drops between rounds and better overall convergence. 
When $K$ is sufficiently large, and clients have non-IID data while training for longer locally, the model experiences fewer convergence issues and tends to achieve performance closer to that of IID data, reaching approximately 95\% to 97\% at the end of the ten epochs.

These results show that, in real-world scenarios, it is crucial to find an optimal set of hyperparameters to better handle the heterogeneous and highly dense data present in this ecosystem.

\begin{figure}[ht!]
\centering
        \includegraphics[width=.4\textwidth]{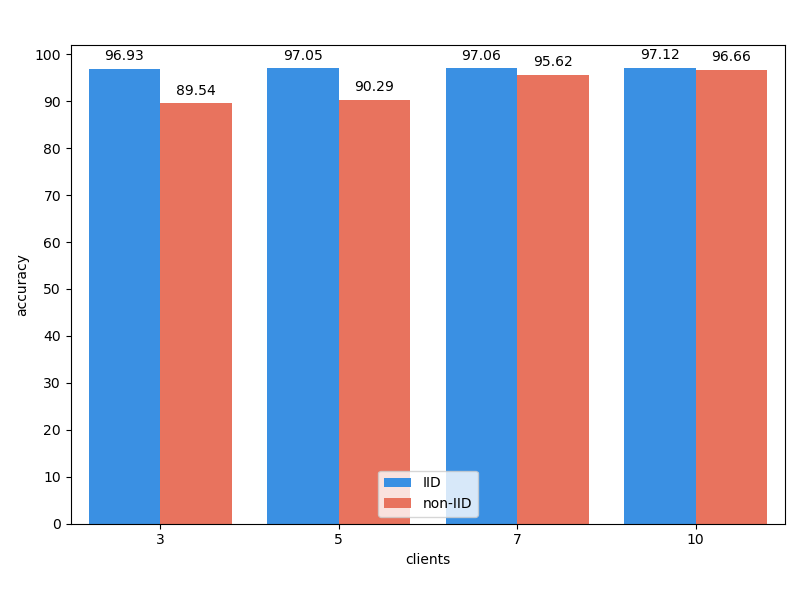}
        \caption{Accuracy by client count per round}
        \label{compAcc}
\end{figure}

The trend in Fig. \ref{compAcc} confirms the hypothesis that the model trained on non-IID data benefits from an increasing number of clients participating in the training round. Even though clients differ from each other, they successfully build a global model using information from all of them. Despite converging to a lower accuracy compared to IID data, the model still demonstrates convergence.

As shown in Table \ref{tab:accuracy_comparison}, while federated learning achieves slightly lower accuracy compared to centralized training, it still reaches a high accuracy of 97.12\%. with IID data versus 98.20\% in the centralized reference, it offers the crucial advantage of enhanced data privacy. Notably, the performance in non-IID scenarios, peaking at 96.66\% accuracy with 10 clients, underscores the robustness of federated learning models under realistic, heterogeneous data distributions, which is critical for deploying machine learning solutions in diverse and decentralized environments. 

Table \ref{tab:inference_time} shows the inference time comparison of our model on different hardware platforms. Our reference is the Raspberry Pi 4 Model B, which has the longest inference time but still demonstrates the model's adaptability and efficiency, even with limited computational resources.

\begin{figure}[ht!]
    \centering
    \includegraphics[width=0.75\linewidth]{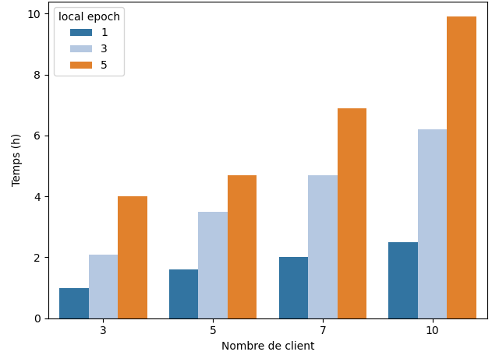}
    \caption{Evaluation of the global model training time with non-IID data}
    \label{fig:federated_training time}
\end{figure}

The Fig. \ref{fig:federated_training time} illustrates the training time (in hours) as a function of the number of clients, with varying configurations for local epochs (1, 3, and 5 epochs per client). As the number of clients increases, the overall training time grows for all local epoch settings. For instance, with 10 clients, the training time is substantially higher compared to the case with only 3 clients. Additionally, for a given number of clients, higher local epochs result in longer training times. The orange bars, representing 5 local epochs, consistently show the longest training time, followed by the light blue bars (3 local epochs) and the dark blue bars (1 local epoch). The growth in training time appears to scale with both the number of clients and the number of local epochs. For example, with 10 clients, the training time for 5 epochs is approximately 9.5 hours, while for 1 epoch, it is just over 3 hours. This figure highlights the trade-off between the number of clients, the number of local epochs, and the training time, suggesting the need for careful selection of these parameters to balance training efficiency and performance.

\begin{table}[ht!]
\centering
\caption{Accuracy comparison of centralized and federated learning approaches}
\label{tab:accuracy_comparison}
\begin{tabular}{@{}lccc@{}}
\toprule
Configuration & Centralized & Federated IID & Federated non-IID \\ 
\midrule
Reference        & \textbf{98.20\%}  & --       & --       \\
Observed         & 97.79\%  & --       & --       \\
3 Clients        & --       & 96.93\%  & 89.54\%  \\
5 Clients        & --       & 97.05\%  & 90.29\%  \\
7 Clients        & --       & 97.06\%  & 95.62\%  \\
10 Clients       & --       & \textbf{97.12}\%  & \textbf{96.66\%}  \\
\bottomrule
\end{tabular}
\end{table}

\begin{table}[ht!]
\centering
\caption{Inference time on different hardware}
\label{tab:inference_time}
\begin{tabular}{|l|r|}
\hline
\textbf{Hardware} & \textbf{Inference Time (s)} \\ \hline
Quadro RTX & 0.00263 \\ \hline
Intel Xeon(R) @ 3.40GHz & 0.03523 \\ \hline
Intel(R) Core(TM) i5 @ 2.40GHz & 0.0874 \\ \hline
Raspberry Pi 4 @ 1.5GHz* & 0.450 \\ \hline
\end{tabular}
\end{table}

\section{CONCLUSION}
The study presented in this paper demonstrates the effectiveness of large language models in cybersecurity, specifically for intrusion detection. We leveraged BERT, an encoder-type LLM, focusing on the upper layers of the model to perform detection tasks due to their focus on syntactic aspects, which is ideal for our task.
We demonstrated that the speed of model convergence is related to the nature of the data distribution. Despite the extreme heterogeneity of data in real 5G ecosystems, it is possible to build a model that provides an effective anomaly-based intrusion detection solution for the devices in that ecosystem. A 5G device can benefit from this model because it is able to effectively identify abnormal traffic.
Future work will explore more suitable aggregation techniques based on the nature of the data. Emphasis will be placed on energy efficiency and detecting attacks based on their frequency, speed, and targets.

\section*{Acknowledgment}
We would like to express our gratitude to the regional computing center ROMEO of the Grand Est region, France, for providing the computational resources necessary for this research.

\bibliographystyle{plain}
\bibliography{bibliography.bib}

\end{document}